\documentclass[12pt,preprint]{aastex}
\usepackage{graphicx}
\usepackage{epstopdf}

\usepackage{natbib} 

\begin{document}

\title{Lack of Transit Timing Variations of OGLE-TR-111b: A re-analysis with six new epochs\altaffilmark{1}}

\author{E. R. Adams\altaffilmark{2},M. L\'opez-Morales\altaffilmark{3}, J. L. Elliot\altaffilmark{2,4},  S. Seager\altaffilmark{2,4},  D. J. Osip\altaffilmark{5}}

\altaffiltext{1}{This paper includes data gathered with the 6.5 meter Magellan Telescopes located at Las Campanas Observatory, Chile.}
\altaffiltext{2}{Department of Earth, Atmospheric, and Planetary Sciences, Massachusetts Institute of Technology, 77 Massachusetts Ave., Cambridge, MA, 02139}
\altaffiltext{3}{Hubble Fellow; Carnegie Institution of Washington, Department of Terrestrial Magnetism, 5241 Broad Branch Road NW, Washington, DC 20015-1305}
\altaffiltext{4}{Department of Physics, Massachusetts Institute of Technology, 77 Massachusetts Ave., Cambridge, MA, 02139}
\altaffiltext{5}{Las Campanas Observatory, Carnegie Observatories, Casilla 601, La Serena, Chile}
\begin{abstract}

We present six new transits of the exoplanet OGLE-TR-111b observed with the Magellan Telescopes in Chile between April 2008 and March 2009.  We combine these new transits with five previously published transit epochs for this planet between 2005 and 2006 to extend the analysis of transit timing variations reported for this system. We derive a new planetary radius value of $1.019 \pm 0.026~R_{J}$, which is intermediate to the previously reported radii of $1.067\pm0.054~R_J$ \citep{Winn2007} and $0.922\pm0.057~R_J$ \citep{Diaz2008}. We also examine the transit timing variation and duration change claims of \citet{Diaz2008}. Our  analysis of all eleven transit epochs does not reveal any points with deviations larger than $2\sigma$, and most points are well within $1\sigma$. Although the transit duration nominally decreases over the four year span of the data, systematic errors in the photometry can account for this result. Therefore, there is no compelling evidence for either a timing or a duration variation in this system. Numerical integrations place an upper limit of about $1~M_{\oplus}$ on the mass of a potential second planet in a 2:1 mean-motion resonance with OGLE-TR-111b.

\end{abstract}


\section{Introduction}

Transiting exoplanets provide a wealth of information for studies of the physical parameters of planets and their environments. For example, the combination of several accurately timed transits of a known transiting exoplanet can be used not only to improve estimates of the planetary radius and orbital parameters of the star-planet system, but also to detect additional objects. Detecting potential variations of parameters such as the inclination and duration of the transits would indicate a precesing planetary orbit, potentially caused by another planet \citep{MiraldaEscude2002}. We can also use transit timing to search for additional planets or moons, as discussed in several recent theory papers \citep{Holman2005, Agol2005, Heyl2007, Ford2007, Simon2007, Kipping2009a, Kipping2009b}. The idea is that the presence of additional objects will perturb the orbit of the transiting planet, producing transit timing variations (TTVs) or transit duration variations (TDVs). Those TTVs and TDVs can be detected by monitoring transits over many orbital periods. The absence of such variations can be also used to place limits on the mass and orbital parameters of additional objects in those planetary systems and to gain insight into the systems' architectures.

Recent observations show hints of timing variations for some transiting planets, but no definitive detection of additional planets or satellites has been reported using this technique. The most interesting results so far are (1) the absence of TTVs in several systems, which do not host planets more massive than several Earth masses in low-order resonant orbits (see a summary of constraints that can be placed in Table~\ref{table:ttvlit}); (2) the tentative detection of TDVs in GJ436, roughly 3 minutes per year \citep{Coughlin2008}, a trend consistent with the presence of a low-mass companion ($<12~M_{\oplus}$) in a close exterior but non-resonant orbit; this result is consistent with the $8~M_{\oplus}$ limit placed by transit timing \citep{Bean2008}; and (3) the preliminary detection of TTVs with a maximum residual of $156 \pm 48$ sec ($3.3\sigma$) over a period of 2 years reported by \citet{Diaz2008} for OGLE-TR-111b, the subject of this paper.
 
OGLE-TR-111b is a $0.5 M_J$ hot Jupiter orbiting its host star, a faint ($I=15.5$) K dwarf, every 4.01 days. This object was first announced as a transiting planet candidate by \citet{Udalski2002}, and was confirmed to have planetary mass by \citet{Pont2004}. The physical parameters of the planet were refined over the next two years, with several new radial velocity measurements \citep{Gallardo2005, Silva2006, Santos2006}. The first high precision transit photometry was provided by \citet{Winn2007}, with two \emph{I}-band transits of the planet on 2006 Feb 21 and Mar 5. Shortly after, \citet{Minniti2007} published a \emph{V}-band transit from 2005 April 9 and noted that the midtime occurred 5 minutes earlier than expected from the ephemeris in \citet{Winn2007}, although with only three epochs they could draw no firm conclusions. A follow-up paper by \citet{Diaz2008} reported two consecutive \emph{I}-band transits of OGLE-TR-111b on 2006 Dec 19 and 23. Combining all five epochs, they concluded that the previously claimed TTVs were real, with the residuals spanning $-156\pm48$ to $+98\pm39$ seconds. Among other scenarios, they noted that if OGLE-TR-111b were in an eccentric orbit with $e\sim0.3$, the observed TTVs would be consistent with the presence of an Earth-mass planet near an exterior 4:1 resonant orbit. Additionally, \citet{Diaz2008} noted two parameters with marginally discrepant values across the five transits (see Table~\ref{table:litparams}). Compared to the results from \citet{Winn2007}, the \citet{Diaz2008} values for the planetary radius disagreed at the 10\% level, or $1.3\sigma$, and the total transit duration differed by $1.6\sigma$. The radius ratio discrepancy was suggested to be the result of the parameters chosen for the image subtraction photometry, which focused on precise timing rather than on an accurate transit depth determination. The duration variation, if real, could be due to a perturber decreasing the orbital inclination, which would offer another way of determining the properties of the third body in the system suggested by their TTVs.

Here we present six new transits observed during 2008 and 2009, which double the number of high-quality transit light curves available for OGLE-TR-111b. In \S~\ref{section:obs} we describe the collection and analysis of the new data. In \S~\ref{section:fitting} we describe the transit model fitting, and discuss additional sources of error not included in the formal fit. In \S~\ref{section:results} we combine the six new transits with the five previously published observations and provide a new analysis of parameter variation in the OGLE-TR-111 system. In \S~\ref{section:conclusions} we discuss the implications of our results.

\section{Observations and Data Analysis}
\label{section:obs}

All six new transits were observed between April 2008 and March 2009 in the Sloan $i'$ filter with the new MagIC-e2v camera\footnote{The MagIC-e2v detector, which shares a dewar with the older SiTe CCD, is identical to the red CCD on HIPO, a fast read-out direct imaging camera and one of the first generation instruments to be flown on SOFIA; both cameras use the LOIS control software \citep{Dunham2004, Taylor2004,Osip2008}.} on Magellan. The MagIC-e2v camera has a field of view of 38\arcsec\ x 38\arcsec\ and a plate scale of 0.037\arcsec\ per pixel unbinned.  With such high resolution and good average seeing at the site, blends are minimized and aperture photometry can be successfully applied even in fairly crowded fields. The camera can be operated in two different modes: single exposure mode, with a readout time of about 5 seconds per exposure, and frame transfer mode, with a readout time of only 3 milliseconds between frames in an image cube.  Our first four transits were observed in single exposure mode. The frame transfer mode first became available after engineering in July 2008, and was used for the last two transits of OGLE-TR-111b. The gain and read noise of the first four transits were 2.4 e-/ADU and 5.5 e- per pixel, respectively; after engineering, these values were changed to the current values of 0.54 e-/ADU and 5 e- per pixel.

The exposure times during each transit were adjusted to maintain a minimum count level of about $10^6$ integrated photons, both for the target and multiple nearby comparison stars. For the 2008 transits, we collected unbinned (1x1) data with exposure times between 30 and 120 sec, depending on the observing conditions, with an additional readout overhead per exposure of 5 seconds per frame. The 2009 data were collected in frame-transfer mode with the camera binned 2x2, which yielded an improved sampling rate of 15-30 sec per frame. Details of the observing settings are noted in Table~\ref{table:obsparams}.

Accurate timing is of the utmost importance for this project, so special care was taken to ensure that the correct times were recorded in the image headers. For the 2008 transits, the start times for each image were recorded from a network time server, which was verified by eye to be synchronized with the observatory's GPS clocks at the beginning of each night. For the 2009 observations, the times came from a PC104 (a small embedded control computer), which received unlabeled GPS pulses every second. As with the network time server, the PC104 was synchronized with the observatory's GPS before each transit observation. In both cases the time signals written to the image headers agree within one second with the GPS time. During both 2009 transits, a software failure caused the times for a few image cubes to not be recorded in the headers of the images, but we were able to reconstruct the observation times with precisions better than a second from detailed system logs. One second is a conservative estimate of the intrinsic error for the start time for each frame, and is significantly smaller than the mid-transit times errors.

\subsection{Data Analysis}

All  data were overscan corrected and flattened using IRAF.\footnote{IRAF is distributed by the National Optical Astronomy Observatories, which are operated by the Association of Universities for Research in Astronomy, Inc., under cooperative agreement with the National Science Foundation.} The photometry was performed using the IRAF routine \emph{phot}, part of the \emph{apphot} package. Depending on the binning applied and the seeing during each transit, a subset of apertures between 6-25 pixels in radius were examined for the target star and each comparison star. Between one and seven stars were selected as comparisons from the list of 10-20 field stars present in each frame. The comparison stars had to pass several selection criteria: to be similar in brightness to the target, to not be blended, and to not be variable.  The best apertures were identified as the ones which yielded the smallest scatter in the out-of-transit flux. The choice of position of the sky background annulus also influenced which stars could be used for comparison; see \S~\ref{section:systematics} for a discussion of systematic errors resulting from aperture settings. We explored sky regions with inner radii from 20-40 pixels and 10-30 pixels width, and selected the one which (a) provided several comparison stars of suitable brightness that were not variable, and (b) produced the lowest noise in the out-of-transit baseline while not introducing spikes or other obvious problems in the transit light curve. 

In each of the resultant light curves, the out-of-transit baseline was examined for linear correlations with several variables: airmass, seeing, telescope azimuth, (x,y) pixel location, and time since beginning of transit. The parameters chosen are either directly correlated with physical phenomena that affect photon rates (e.g. seeing, airmass), or are proxies for other effects (e.g. the telescope azimuth is correlated with the de-rotator rates, which were not recorded for several transits). Only the 2009 transits exhibited significant trends, against telescope azimuth (both) and seeing (20090313). For each transit, we corrected those trends by successively subtracting linear fits to each variable. Detrending was critical in producing usable light curves for these two transits, but may also have introduced smaller-order systematic effects, particularly in the transit depth (see also \S~\ref{section:systematics}).

We now briefly describe the observations and the photometric reduction of each transit dataset. All transit fluxes and times are available online; an excerpt is shown in Table~\ref{table:data}.

\subsubsection{20080418}

Transit 20080418 was observed during engineering time just after the e2v CCD was first installed on the telescope; due to engineering constraints, only the second half of the transit was observed. The field was repositioned before egress to eliminate diffraction spikes from a nearby bright star by moving it further off-chip. The airmass was low and fairly constant (1.2-1.3) and the seeing was good, increasing slightly from 0.4\arcsec\ to 0.5\arcsec\ during the transit. The out-of-transit data showed no apparent trends with the parameters checked.

\subsubsection{20080422}

Transit 20080422 was observed on an intermittently cloudy night with highly variable transparency, with counts on the target star varying by a factor of 6 within a few frames. We found that eliminating the lowest count frames, those with fewer than 300,000 counts on the target star, significantly decreased the scatter of the light curve. The seeing ranged from 0.5-0.6\arcsec, and the airmass was low and fairly constant (1.2-1.3). The out-of-transit data showed no apparent trends with the parameters checked.

\subsubsection{20080512}

Transit 20080512 had stable photometric conditions for the entire pre-transit baseline. During the transit there were two drops in target counts (by a factor of 2) that coincided with sudden seeing jumps (0.4\arcsec\ spiking to 0.6\arcsec). The field also drifted substantially (by $\sim$100 rows and $\sim$100 columns) due to tracking problems; about 30 minutes of post-transit baseline had to be discarded because of strong image elongation. The airmass ranged from 1.2 to 1.7. The out-of-transit data showed no apparent trends with the parameters checked.

\subsubsection{20080516}

Transit 20080516 had very stable photometric conditions for most of the transit. The seeing gradually increased from 0.4\arcsec\ to 0.6\arcsec, and the airmass ranged from 1.2 to 2.0. The star also drifted substantially toward the end of transit (by $\sim200$ rows and $\sim200$ columns) for unknown reasons. The out-of-transit data showed no apparent trends with the parameters checked.

\subsubsection{20090217}

Transit 20090217 was the first of OGLE-TR-111b to be observed with the new frame transfer mode. The seeing fluctuated from 0.7-1.1\arcsec, while the airmass decreased from 1.8 to 1.2. This transit was detrended for a slope correlated with the telescope azimuth.

\subsubsection{20090313}

Observations for transit 20090313 began late due to telescope problems and thus there is no pre-transit baseline. Seeing conditions were initially very poor, spiking to 2\arcsec, but improved substantially during the second half of the transit, to around 0.6\arcsec. The airmass ranged from 1.2 to 1.5. The huge disparity in seeing made photometry on this transit challenging, and the best light curve resulted from using a small aperture around the target and only the brightest single comparison star, which saturated for a few frames that had to be discarded. This transit was detrended for slopes correlated with the seeing and the telescope azimuth.

\subsubsection{Literature light curves}

To eliminate any uncertainty in comparing our transits to previously published transits, which might result from different models, fitting procedures, or fixed parameter values, we have obtained tables of the times and fluxes for each of the five transits drawn from the literature:  20050409 \citep{Minniti2007}, 20060221 and 20060305 \citep{Winn2007}, and 20061219 and 20061223 \citep{Diaz2008}. We have used the original photometry, except for converting from magnitudes into fluxes (where applicable) and converting the mid-exposure times from Heliocentric Julian Day, or HJD, into Barycentric Julian Day, or BJD, to be consistent with the rest of our analysis; the difference between the two time systems is much smaller than our errors, though, a few seconds at most. All values presented for the literature light curves are taken from our re-analysis of the published photometry using our model and fitting, unless otherwise noted.

\begin{figure}
\includegraphics*[scale=0.5]{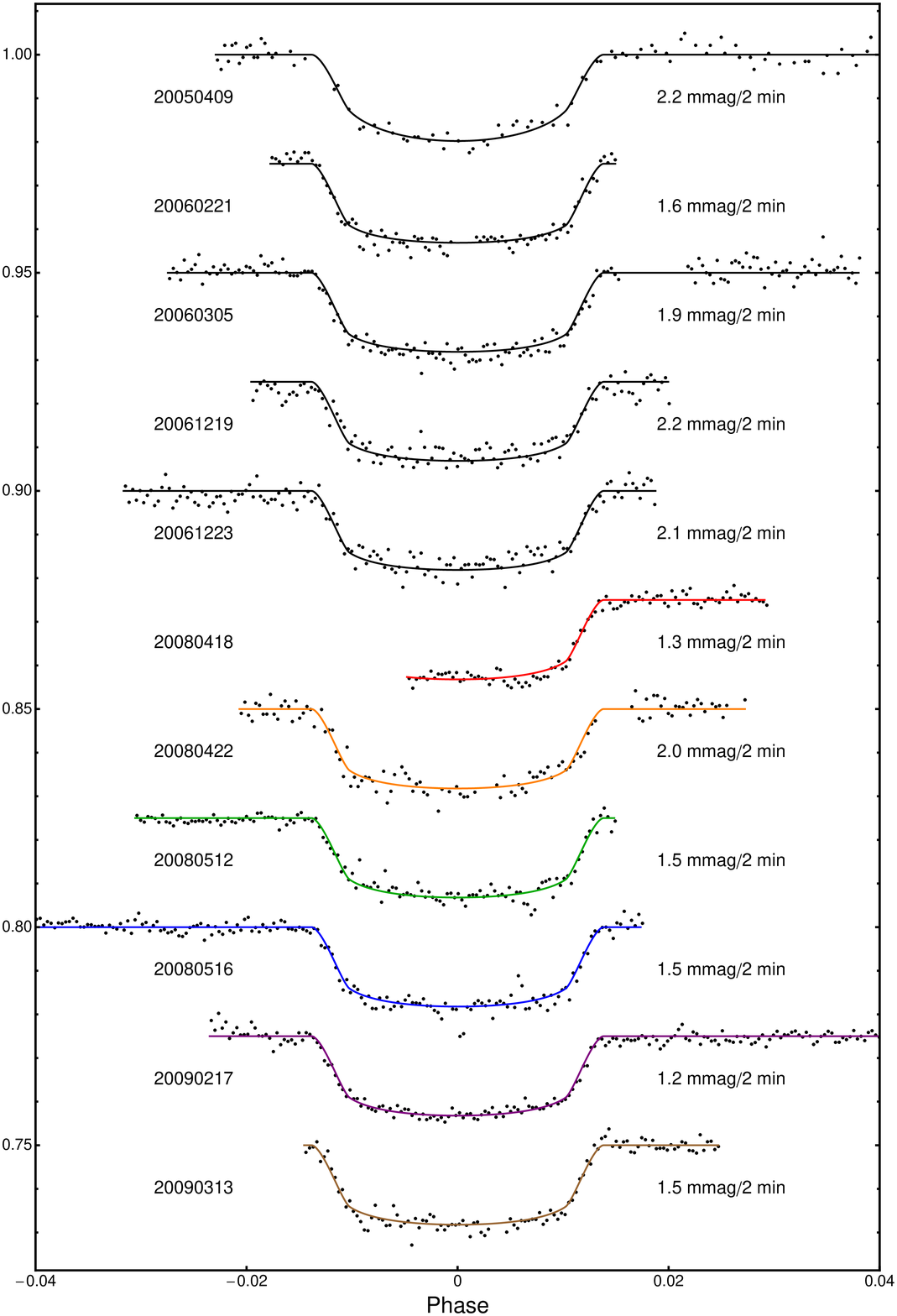}
\caption{Eleven transits of OGLE-TR-111b. All available high-quality light curves are plotted vs. orbital phase, with the data binned to 2 minutes to aid comparison. The joint model fit (solid lines) were calculated using the parameter values in Table~\ref{table:mcmc}; the stated standard deviation is the residuals from the joint model fit. Table~\ref{table:data} shows the unbinned data; a full table is provided online.
\label{fig:transits}}
\end{figure}

The light curves for all six new transits and the five literature transits are shown in Figure~\ref{fig:transits}, together with the best joint model fit, which will be described in detail in \S~\ref{section:fitting}.

\section{Transit fitting results}
\label{section:fitting}

\subsection{Model}

Each light curve was fitted with the \citet{Mandel2002} algorithm to generate analytical models, using the basic optimized model-fitting code described in \citet{Carter2009}, but without the wavelet analysis. In the models we assumed that OGLE-TR-111b has zero obliquity, oblateness and orbital eccentricity. We used a quadratic limb darkening law of  the form
\begin{equation}
I(r) = 1 - u_1 (1 - \sqrt{1-r^2}) - u_2 (1  - \sqrt{1-r^2})^2,
\end{equation}
with the initial parameters for $u_1$ and $u_2$ set to the values for the appropriate filter \citep{Claret2000,Claret2004}. Although analyses of different limb darkening laws have shown that using a non-linear limb darkening law is important \citep{Southworth2008}, it is generally not possible to fit both quadratic limb-darkening coefficients except on the highest quality, typically space-based, data. We thus fixed the quadratic term $u_2$ and only fit for the linear term, $u_1$. (We also fixed $u_1$ for the sparsely-sampled transit on 20050409, the only one observed in $V$ band.) The values for $u_1$ and $u_2$ are calculated using the \emph{jktld} program by \citet{Southworth2008}\footnote{http://www.astro.keele.ac.uk/~jkt/codes/jktld.html}, assuming $T=5044~K$, $\log{g}=4.25~\textrm{cm/s}^2$, $[M/H] = 0$, and $V_{micro} = 2$~km/s. We used the limb darkening values corresponding to the Sloan $i'$ filter for the new data, the $V_C$ filter for the transit from \citet{Minniti2007}, and the $I_C$ filter for the transits from \citet{Winn2007} and \citet{Diaz2008}, as listed in Table~\ref{table:mcmc} (though we note for completeness that the actual filters used in the literature light curves, Mould \emph{V}, CTIO \emph{I} and Bessel \emph{I}, respectively, do not correspond precisely to the Cousins \emph{V} and \emph{I} filter parameters that were available). We fixed the orbital period to $P=4.01445$ days; later experiments with slightly different values had little effect. The other free parameters in the model are the radius ratio, $k$, inclination, $i$, semimajor axis in stellar radii, $a/R_*$, out-of-transit flux, $F_{OOT}$ and transit midtime, $T_C$.  We assume throughout the fits that $M_*=0.81~M_{\odot}$, $R_*=0.83~R_{\odot}$, and $M_{p}=0.52~M_J$, based on the spectroscopic work of \citet{Santos2006}.  

\subsection{Light curve fits}

\begin{figure}
\includegraphics*[scale=0.35]{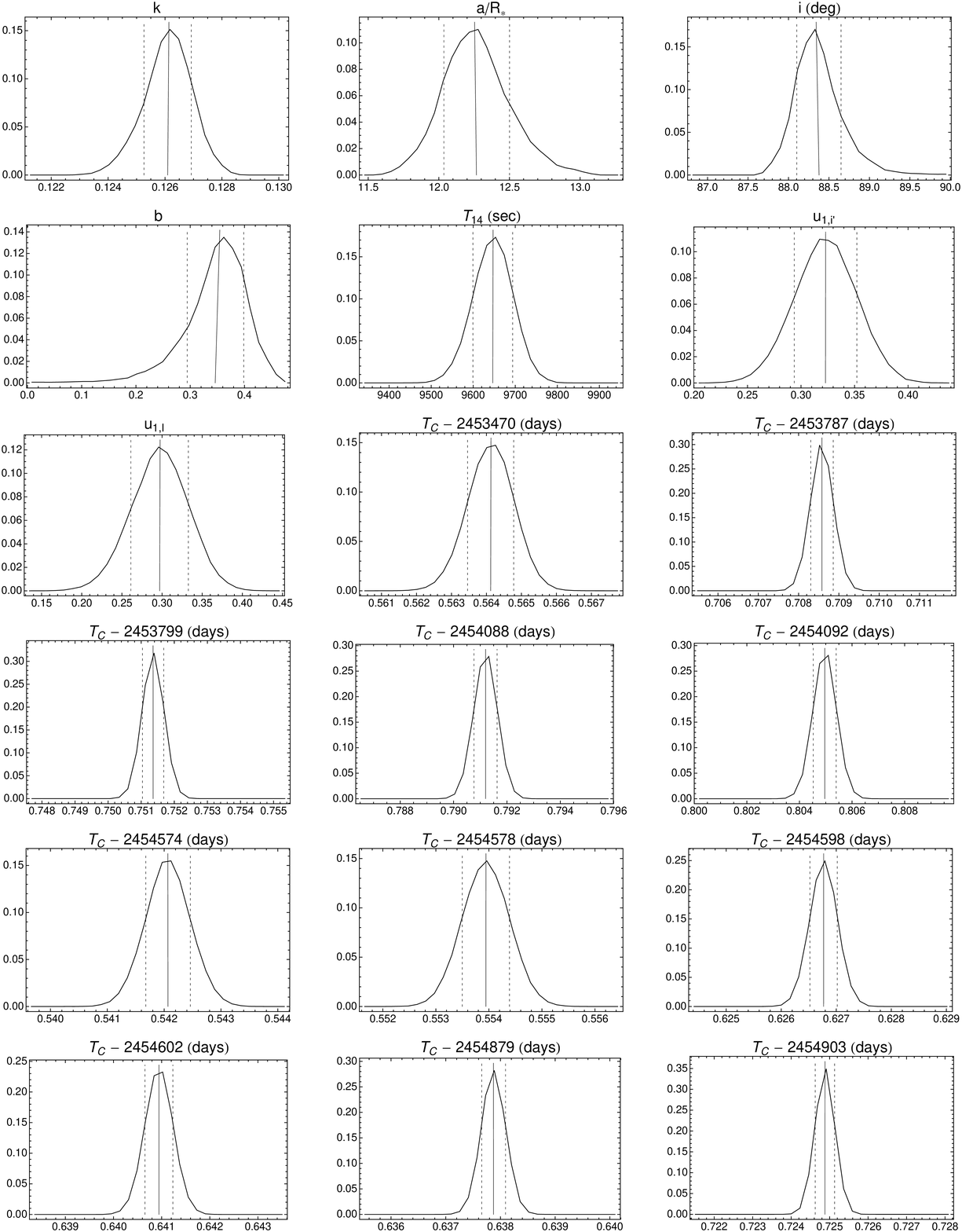}
\caption{Smoothed histogram of normalized parameter distributions, from which the parameters in Table~\ref{table:mcmc} are derived. The solid line is the median value (which is very close to the mean value in all cases). Note that several of the distributions, particularly $a/R_*$, $i$, and $b$, are not strictly Gaussian. The dashed lines show the 68.3\% credible interval. These values were calculated for 2,850,000 links. 
\label{fig:dist}}
\end{figure}

To determine the best fit value and error of each model parameter, we used a Monte Carlo Markov Chain (MCMC) method, as described in \citet{Carter2009}. The initial values for each parameter were computed by a joint least squares fit to each light curve independently. We then weighted each light curve by the reduced $\chi^2$ of this fit so that the new reduced $\chi^2=1$. (By doing this we are assuming the transit model is correct in order to determine the error on each transit, rather than assuming a noise model, e.g. photon noise, in order to test the transit model.) A joint least squares fit of all weighted transits had a reduced $\chi^2=1.05$. Starting from the initial least squares values, we constructed chains of 1,000,000 links, where the acceptance rate for each parameter is between 20-60\%.  We fit all eleven transits simultaneously, assuming common values for $k$, $i$, $a/R_*$, and $u_{1,x}$ (where $x$ is the appropriate filter), with $u_{2,x}$ fixed, but fitting each transit for its own $F_{OOT}$ and $T_C$. We did not fit for an airmass slope (see \S~\ref{section:systematics}. The first 50,000 points of the MCMC were discarded to eliminate bias toward the initial conditions. We created three independent MCMC chains, checked that the Gelman-Rubin statistic \citep{Gelman1992} is close to 1 to ensure convergence, and then combined the chains to determine the distribution of all parameters, including the total duration of the transit, $T_{14}$ (the time from first to fourth, or final, contact), and the impact parameter, $b$, which are derived from fitted parameters. We plot the best model fit with the data in Figure~\ref{fig:transits} and tabulate the fit results in Table \ref{table:mcmc}, where we report for each parameter the median value and the 68.3\% credible interval (the equivalent to a $1\sigma$ standard deviation if the distribution is Gaussian). The distributions for each parameter are shown in Figure~\ref{fig:dist}.

The new radius ratio for OGLE-TR-111b based on an analysis of all eleven light curves yields a planetary radius $R_p=1.019\pm0.007~R_J$, if we consider only the formal fit errors; accounting for the error on the stellar radius, which is now the dominant source of error, we find a more realistic error bar is $R_p=1.019\pm0.026~R_J$. Note that if we only use the six new light curves, which have more consistent radii, the value for the radius ratio is very similar (formal fit $R_p=1.015\pm0.009~R_J$, or $R_p=1.015\pm0.026~R_J$ with stellar errors).

As a test of the robustness of our parameter determination, we also ran additional MCMC fits for each transit independently, with results in Table~\ref{table:indfits}. Although most of the parameters agree within the formal $1\sigma$ errors between the individual and joint fits, there are some notable exceptions. In \S~\ref{section:variations} we investigate variability (previously noted by \cite{Diaz2008}) in both the radius ratio, $k$, and the total transit duration, $T_{14}$.

\subsection{Systematic Errors and Correlated Noise}
\label{section:systematics}
One of the most apparent results from the fits to individual light curves was that the radius ratios are similar for transits observed on the same instrument and reduced by the same group. This may indicate a degree of subjectivity in the light curve generation process, both from the choice of photometry method (e.g. aperture, image subtraction, deconvolution, etc.) and from the specific choice of reduction parameters (e.g. aperture size and sky region for aperture photometry). These choices can result in systematic errors in the transit depth, particularly when comparing transits from multiple sources. Both image subtraction and aperture photometry require fine-tuning a number of parameters, and there is no single prescription for how to get the absolute best light curve: the same method applied to the same transit could produce similar quality light curves, as measured by the scatter of residuals or out-of-transit flux, which nonetheless differ in depth by more than the formal fitted errors. It has been noted by \citet{Winn2007} that with image subtraction, slight changes in both the difference flux and the reference flux can cause the measured radius ratio to vary by a few percent of its value, although their estimate of that effect on their own data, $\delta k =0.0002$, is much less than our formal fit error of $0.0008$. This effect was also alluded to by \citet{Diaz2008} as an explanation for their shallow depths compared to previous results, although they did not provide a numerical estimate of the magnitude of this effect.  An analysis by \citet{Gillon2007} of a different transiting planet, OGLE-TR-132b, found that image subtraction is particularly prone to misestimating the transit depth, compared to the alternative methods of aperture and deconvolution photometry. This effect for OGLE-TR-132b in their data causes the radius ratio to differ by 1-2\% depending on the choice of parameters. If similar levels of error were present for OGLE-TR-111b, particularly in the \citet{Diaz2008} curves which were acknowledged to not be optimized toward finding the correct depths, the systematic error on the radius ratio would be $0.0013-0.0025$, comparable to the formal fit error of $0.002$ on the individual curves. Assuming a median value of $0.002$ for the systematic error and adding it in quadrature with the formal fit error, a better error estimate on the radius ratio would be $0.003$. 

We attempted to quantify the systematic error for aperture light curve generation as follows. For each of our light curves, we used 4-5 sets of apertures, sky radii and widths, and different comparison stars, with a goal toward minimizing the scatter in the out-of-transit baseline. The choice of comparison stars in particular depends on the aperture and sky choices, because certain stars are usable under some choices but not others, especially stars of very different brightnesses. Additionally, we examined the effect of detrending light curves, which must be done carefully because systematic trends in the data can distort the measured radius, but so can an incorrectly-removed slope. We only detrended transits which had strong slopes in the out of transit baseline or otherwise had distorted shapes, and then only for the parameters which best corrected the shape defects. It is possible that residual trends against parameters we did not consider, or slightly nonlinear trends, could remain in the data and distort our estimate of the radius; we chose to stick with linear trends against a few meaningful physical parameters to avoid introducing unnecessary complexity. To test whether any important slopes remained, we ran a joint MCMC fit which includes a differential extinction term (a trend with airmass), but found that the fitted airmass slopes were slight and the difference in resulting parameters were in all cases less than the formal $1\sigma$ errors. The results of our explorations of both the aperture and sky choices and the detrending parameters are that we can produce light curves with similar shapes and scatter, and that the radius ratios vary by 0.001-0.004 for an individual transit. (Some transits are much more resilient to parameter choices than others.) Thus, for aperture photometry also the systematic error on the radius ratio based solely on the parameter choice is of order the formal fit errors. 

To estimate the amount of correlated noise in the light curves, we used two methods. The first is residual permutation, which shifts the residuals for each transit through every point in time and adds it to the best model fit; we also assumed time invariance and reversed the residuals, then permuted again, for a total of 214-1600 curves for each transit. We fit a least-squares transit model to each permuted curve for all eleven transits.  For the radius ratio and the transit midtime, the errors from the residual permutation method for both values were greater by a factor of 1-3, depending on the light curve (note that this factor is not the same for each parameter; see Table~\ref{table:indfits}). We got similar results when we ran a joint fit of all 11 transits with 10,000 curve ensembles, randomly selecting for each transit one of its individually permuted light curves. 

An alternative way to estimate the error contributed by correlated noise is to calculate how the noise scales with time averaging \citep{Pont2006}. We calculate the standard deviation on the residuals in bin sizes from 10-30 minutes and compare that value to what we would expect if the noise behaved like Poisson noise (i.e., a decrease in the noise with $\sqrt N$ points). We calculate the amount by which the real noise is greater than the estimated noise, and find that it is greater by a factor of 1.5-3 times the purely Poisson noise level, depending on the transit. The increased noise factors agree with the values found by residual permutation, and for simplicity we use the scaled errors from the residual permutation throughout.

For all of these reasons, the formal fit errors reported in Table~\ref{table:mcmc} are underestimated due to both systematic errors and correlated noise. The errors we adopted have been inflated based on the residual permutation method. This means that, once systematics are included, neither the observed timing variations for 20050409 and 20080418 nor the duration variation are statistically significant (see \S~\ref{section:timing} and \S~\ref{section:variations}).

\section{Results}
\label{section:results}

\subsection{Timing}
\label{section:timing}

\begin{figure}
\includegraphics*[scale=0.5]{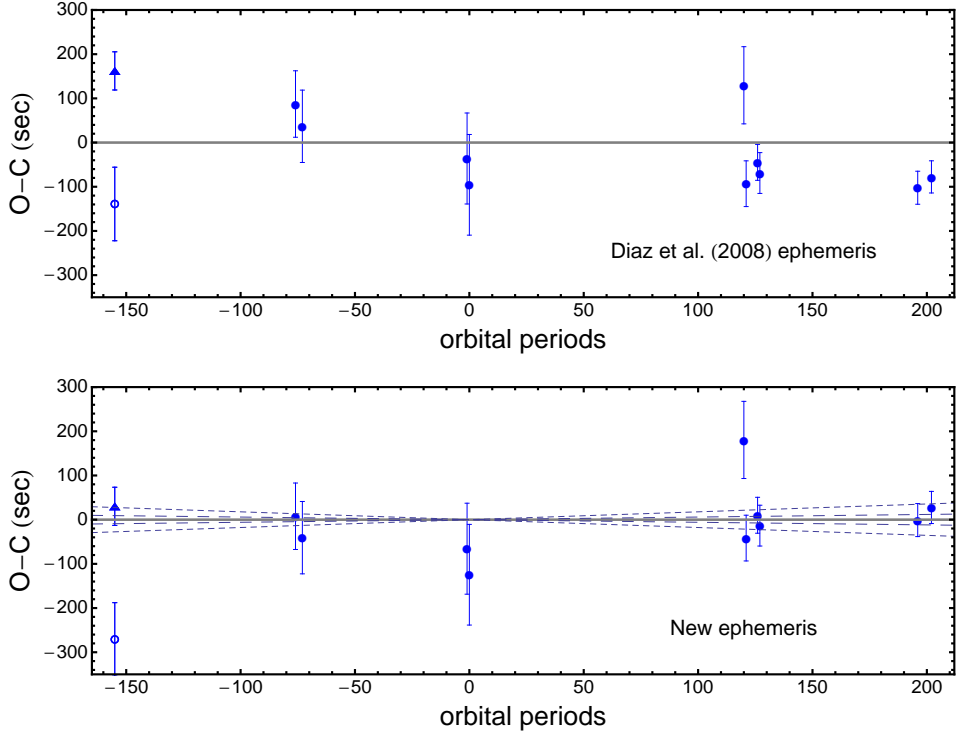}
\caption{Observed minus calculated midtimes for OGLE-TR-111b. Top panel: timing residuals for eleven transits using the ephemeris from \citet{Diaz2008}. Bottom panel: timing residuals using the new ephemeris (Equation~\ref{eqn1}. The O-C values and errors are shown in Table~\ref{table:ominuscvalues}, and were calculated using the formal fit midtimes reported in Table~\ref{table:mcmc} but rescaling the errors to more realistically account for systematic noise (see \S~\ref{section:systematics}). The solid line represents zero deviation from expected time of transit, while the dashed lines represent the $1\sigma$ and $3\sigma$ errors on the calculated orbital period, indicating the slopes that result for a mis-determined period. We plot our calculated midtime for 20050409, based on the photometry from \citet{Minniti2007}, as an open circle, and the revised midtime and error reported by \citet{Pietrukowicz2010} as a solid triangle. The new ephemeris was calculated using only the solid symbols, i.e., using the \citet{Pietrukowicz2010} time.
\label{fig:ominusc}}
\end{figure}

The central midtimes for all 11 transits that we fit are summarized in Table~\ref{table:mcmc} and illustrated in Figure~\ref{fig:ominusc}. Recently \citet{Pietrukowicz2010} have reanalyzed the photometry for 30 transits of OGLE planets and planet candidates, among them OGLE-TR-111b, and they have found a different midtime than originally reported in \citet{Minniti2007}:  $T_{C,new} = 2453470.5676 \pm 0.0005$, compared to the original published value of  $T_{C,orig} = 2453470.56413 \pm 0.00067$, a difference of 300 seconds. (It is not clear what is the source of such a large shift, but one possibility is a mistake in the UTC-BJD correction.) Significantly, the new time is much closer to the expected time of transit.  

Another potential pitfall when comparing times from multiple groups has recently noted by Eastman and Agol (in prep).  Most researchers, and indeed most common conversion tools (e.g. \emph{barycen.pro} in IDL and \emph{setjd} in IRAF) by default omit the correction from UTC to TT, which in 2009 was 66.184 seconds. We have confirmed that the times published by \citet{Pietrukowicz2010} and \citet{Winn2007} do not account for the UTC-TT correction (personal communications), and we assume that the times in \citet{Minniti2007} and \citet{Diaz2008} likely did not either. We have therefore added the appropriate correction to the reported BJD times for these light curves. (Note that the smaller order deviations introduced by using UTC rather than TT times in calculating the BJD correction terms are at most a few seconds, and for this work those deviations fall well within the timing errors; however, with higher precision data on other systems it would be very important to consistently calculate the BJD times.) All of the transit midtimes in Table~\ref{table:mcmc} have been corrected to the BJD-TT system; additionally, we have added 64.184 seconds to the \citet{Pietrukowicz2010} midtime to get $T_C = 245470.56834$ (BJD), the value used in all subsequent calculation.

The top panel of Figure~\ref{fig:ominusc} shows the ephemeris from \citet{Diaz2008}, derived from the first five transits, along with the timing residuals for all eleven transit based on our joint-fit values (Table~\ref{table:mcmc}). To correct for the linear drift in the residuals, we calculate a new constant-period transit ephemeris, omitting the half-transit (20080418), which has very large errors, and using the new \citet{Pietrukowicz2010} time for 20050409, and find:

\begin{equation}
T_{C} = 2454092.80717(16) [BJD] + 4.0144463(10) N.
\label{eqn1}
\end{equation}
where $T_C$ is the predicted central time of a transit, and $N$ is the number of periods since the reference midtime, and the values in parentheses are errors on the last digits. We find almost identical ephemeris values if we use all transits and if we use the original time for 20050409, although the errors are several times greater in both cases. Our adopted fit has a reduced $\chi^2=0.5$.

The lower panel in Figure~\ref{fig:ominusc} shows the new ephemeris and the timing residuals. (Note that the same times are used in both panels, and only the ephemeris has changed.) With $1\sigma$ errors ranging from 36 s to 114 s, only the original 20050409 time is more than $2\sigma$ from zero, and of the other transits only the half transit 20080418, which is inherently less trustworthy, is more than $1\sigma$. Thus, we conclude that the timing deviations reported by \citet{Diaz2008}, which depended heavily on the old time for 20050409, do not exist, and we see no evidence for timing variations in our data. 

\subsection{Parameter variation}
\label{section:variations}

\begin{figure}
\includegraphics*[scale=0.4]{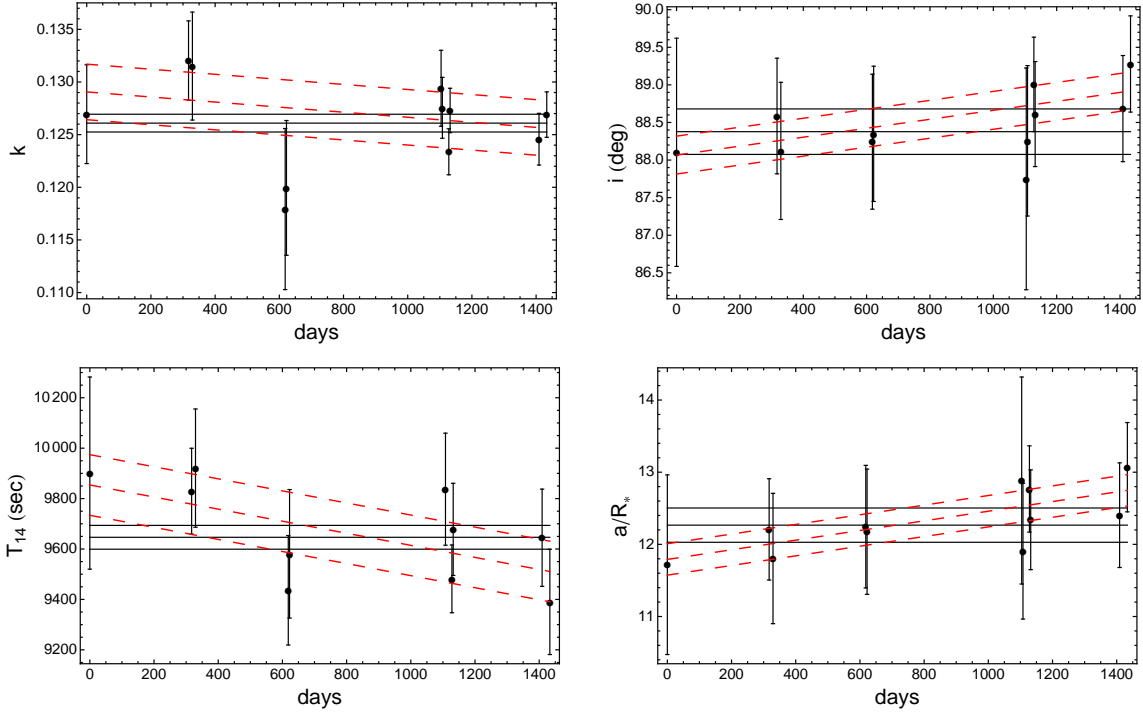}
\caption{Parameter variation vs. time for all transits of OGLE-TR-111b, based on individual MCMC fits (Table~\ref{table:indfits}), for (clockwise, from top-left): $k$, $i$, $a/R_*$, and $T_{14}$. The errors have been scaled upward based on the factor calculated from residual permutation. The values derived from the joint MCMC fit to all transits (Table~\ref{table:mcmc}) are plotted as solid black lines with $\pm1\sigma$ errors. The dashed red lines indicate the best sloped fit with $\pm1\sigma$ errors, although all fits are only marginally significant (within $1\sigma$ of a constant value for the radius ratio and within $2\sigma$ for the other parameters). 
\label{fig:paramvar}}
\end{figure}

\citet{Diaz2008} found that their value for the total duration was 4.4 minutes shorter than that found by \citet{Winn2007}, a $1.6\sigma$ result given the respective quoted errors.  If this decrease is real, it would be of great interest, since a likely explanation would be that the inclination of OGLE-TR-111b is precessing, possibly due to the presence of another planet.  On the other hand, the variation could be due to errors in the photometry or undetected correlated noise. Our values for the best fit duration for each transit are plotted in Figure~\ref{fig:paramvar} and tabulated in Table~\ref{table:indfits}, along with the parameters the duration was derived from: $i$, $k$, and $a/R_*$. Note that the errors in this table have been increased from the formal fit errors by a factor derived from the residual permutation method discussed in \S~\ref{section:systematics}. 

At first glance there does appear to be a decrease in duration over time. To compare the durations of the 10 full transits (excluding 20080418) we fit the data using two models: (1) a flat line, corresponding to a constant duration, with reduced $\chi^2 =0.9$, and a value of $9636 \pm 58$ s, very similar to the joint-transit fit value, and (2) a sloped line, with reduced $\chi^2 =0.6$, and a slope of $-0.24 \pm 0.12$ seconds per day. This is a much shallower slope than the hint of a trend reported by \citet{Diaz2008}, and is not significant given the good fit achieved with the flat line. 

Because the duration is a derived quantity, we must examine the parameters on which it depends ($i$, $a/R_*$, and $k$). Both the inclination and the semi-major axis exhibit slight slopes, but with low significance given the errors ($i$ changes by $0.22\pm0.09$ degrees per year with reduced $\chi^2=0.2$, and $a/R_*$ by $0.0007\pm0.0002$ stellar radii per year with reduced $\chi^2=0.2$). If the duration were really decreasing because the planet is precessing, this would be due to the planet moving away from the center of the stellar disk and hence the inclination would decrease; instead, the best-fit slope for the inclination is slightly positive, another indication that we are not picking up on a real effect. 

On the other hand, the radius ratio does have real variations between transits, given the current photometry. We find variation from a low of $k=0.118\pm0.002$ \citep[transit 20061219, from][]{Diaz2008} to a high of $k=0.132\pm0.002$ \citep[20060221 and 20060305, both from][]{Winn2007}, with the rest of the transits in between (see Table~\ref{table:indfits}). (Note that our individual fits for the radius and error of the two transits from \citet{Winn2007} agree with those cited in that paper, indicating that our fitting methods are comparable.) The best fit line with a slope is not statistically significant (within $1\sigma$ of 0), and the most likely explanation of the variation in radius depths is due to systematic effects in how the photometry was reduced (see \S~\ref{section:systematics}).

If the star were active, the presence of stellar spots or active regions can affect the observed radius depths (see e.g. \cite{Pont2008} for HD189733, \cite{Rabus2009} for TrES-1, and \cite{Czesla2009,Huber2009,SilvaValio2010} for CoRoT-2, all of which are known or theorized to be spotty stars). There is no record of variability for this star in the current literature. We examined two published data sets with observations over multiple non-transit nights: 4 nights of VIMOS data \citep{Pietrukowicz2010,Minniti2007} and data spanning 115 nights from the OGLE survey \citet{Udalski2002}. We found that the long-term flux was stable to within a few mmag in both datasets (3 mmag and 5 mmag respectively). 

Since the trends in the parameters $i$, $a/R_*$, and $k$ are slight and unlikely to be physical, we cannot conclude based on the available data that the observed duration variation is a real effect.

\subsection{Limits on perturber mass}

Although there is no clear evidence of TTVs beyond the $2\sigma$ level in the current dataset, we can use the TTVs reported in Table~\ref{table:ominuscvalues} and shown in Figure~\ref{fig:ominusc} to place upper limits on the mass and orbital separation of a hypothetical perturbing planet in the system. For that purpose we use an implementation of the algorithm presented in \citet{Steffen2005}, kindly provided by D. Fabrycky.

We explored the full perturber's mass parameter space for interior orbits and exterior exterior orbits for orbital periods from 0.9-17.5 days (0.2-4.4 times the orbital period of OGLE-TR-111b), and small initial eccentricity $e_c=0.05$ (we also examined $e_c=0$ and $e_c=0.3$). All the orbits were assumed coplanar and the orbital instability regions in each case were determined following \citet{Barnes2006}. The orbital period of the perturber was increased by a factor of 1.1 for each step. For each period, the mass of the perturber also increased from an initial value of 0.0001 $M_{\oplus}$ until reaching a mass that would produce a TTV equivalent to the $3\sigma$ confidence level of our results (i.e. $\Delta \chi^2 = 9$). We used the time reported by \citet{Pietrukowicz2010} for the transit 20050409 instead of the value we fit from the original photometry from \citet{Minniti2007}, since the revised value is more in line with expectations; simulations run omitting that transit yield similar results. The mass limits placed by these tests are illustrated in Figure~\ref{fig:ominusc}. 

The constraints placed on the perturber's mass are strongest near the low-order mean motion resonances, particularly in the interior and exterior 2:1 resonances, where we are sensitive to objects as small as $1~M_{\oplus}$ and $0.6~M_{\oplus}$, respectively if $e_c=0.05$ ($e_c=0$ yields even smaller constraints, but we choose to cite the more conservative value). No meaningful constraint can be placed on the region of the 4:1 external resonance, which was identified by \citet{Diaz2008} as a possible location for a $1~M_{\oplus}$ perturber that could explain their TTV, but since we do not reproduce their TTV such a perturber is no longer necessary.

Finally, our O-C data, with a $1\sigma$ precision of 36-114 seconds (after accounting for systematic errors), cannot constrain the presence of moons around OGLE-TR-111b, which would introduce a TTV signal of 4.6 seconds for a $1~M_{\oplus}$ moon \citep{Kipping2009b}.

\section{Conclusions}
\label{section:conclusions}

We have tested the previously claimed presence of TTVs and TDVs in the OGLE-TR-111 system by adding six new transit epochs, observed between 2008 and 2009, to the five previously published results by \citet{Winn2007,Minniti2007,Diaz2008}. This new analysis not only doubles the number of available data points, but also extends the TTV baseline from two to four years. In addition, combining the six new transits data allows us to provide a new, more precise value of the radius of this planet. We find a new radius for the planet of  $1.019 \pm 0.026~R_{J}$, which is intermediate to the previously reported radii by \citet{Winn2007} and \citet{Diaz2008}, and is more precise.

We find a slight variation over time of the duration of the transits of OGLE-TR-111b, as well as variations of other parameters, such as the inclination and semimajor axis of the orbit. Those variations could, in principle, be attributed to perturbations of the orbit of OGLE-TR-111b produced through interaction with additional planet(s) in the system, but we demonstrate that the variations can be instead explained by systematic errors in the data, and therefore should not be attributed to other planets.

We have also computed the transit midtimes of our new transits with formal precisions of 20-40 seconds, and more accurate precisions of 35-50 seconds for the full transits (and almost 2 minutes for the half-transit) once systematic errors are considered. The errors on the literature transits similarly increased when the photometry is refit using the same method to account for systematics, to 60-110 seconds depending on the light curve.

A longer time baseline and more precise timing data is still necessary to test further for the presence of other planets in the OGLE-TR-111 system, especially in potentially stable non-resonant orbits, but with the present results we conclude that OGLE-TR-111 belongs in the category of systems summarized in Table~\ref{table:ttvlit} for which there is no sign of additional planets more massive than a few $M_{\oplus}$ in low-order resonant orbits, including a limit of $1~M_{\oplus}$ near the 2:1 resonances. The presence of massive (Earth-like) moons around OGLE-TR-111b is still possible, but to detect those we would require timing precision of a few seconds or better, beyond the current capability of ground-based instrumentation for this system.

\acknowledgements

E.R.A. received support from NASA Origins grant NNX07AN63G. M.L.M. acknowledges support from NASA through Hubble Fellowship grant HF-01210.01-A/HF-51233.01 awarded by the STScI, which is operated by the AURA, Inc. for NASA, under contract NAS5-26555. We would like to thank Brian Taylor and Paul Schechter for their tireless instrument support. We thank the Magellan staff, in particular telescope operators Jorge Araya, Mauricio Martinez, Hern\`an Nu\~nez, Hugo Rivera, Geraldo Valladares, and Sergio Vera, for making these observations possible.  We also thank Josh Carter and Dan Fabrycky for providing software, and Josh Winn for some helpful discussions.

\clearpage

\bibliographystyle{astron}
\bibliography{main}

\newpage

\clearpage

\clearpage

\clearpage

\begin{figure}
\includegraphics*[scale=0.5]{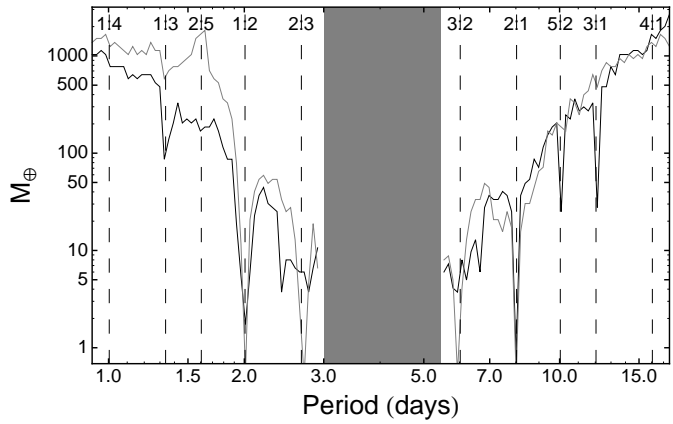}
\caption{Upper mass limit on potential companion planets vs. period of the perturber. We examine companions with initial $e_c=0.05$(black) and $e_c=0.0$ (gray). The constraints are strongest near the 2:1 mean-motion resonance, where objects as small as $1~M_{\oplus}$ would have been detectable; other interior and exterior resonances are also labeled. The shaded grey region shows the instability region for a $1~M_{\oplus}$, following \citet{Barnes2006}. 
\label{fig:maxmass}}
\end{figure}

\clearpage

\begin{deluxetable}{l c  |  lll  |  llr  |  p{10pc} }
\tabletypesize{\scriptsize}
\tablecaption{Maximum Allowed Perturber Mass ($M_{\oplus}$) Constrained by Transit Timing}
\tablehead{System	&Max O-C	 &1:2\tablenotemark{a} & 2:3 & Interior & 3:2 &2:1 & Exterior & Reference\\
 & (sec)		 &                         &        & (non-resonant)& &  & (non resonant) & } 
\startdata
CoRoT-1		&$78\pm24$	& unstable & unstable & unstable  & --     & 4  & 200 ($P<4d$)  & \cite{Bean2009} \\
			&$<60$		& --	 & --  &-- 	 &--       & 1      & 10 ($P<3.5d$),  & \cite{Csizmadia2009} \\
			&			&  	 & 	 &  	 &          &         & 100 ($P<5d$)    &  \\
GJ 436 		&$<60$		& -- 	 & --   & --   & --      & --      & 8 ($P<5 d$)       & \cite{Bean2008}  \\
			&$<120$		& 10	 & 2   & --    & 2      & 20    & 20 ($P<5 d$)     & \cite{Ballard2009}  \\
HAT-P-3 		&$<60$	         & 0.7-1.0 & --   & --     & --     & 30-40  &  --             & \cite{Gibson2009b}  \\
HD 189733 	&$<45 $		& 4  	 & 1    & --   & 8      & 20    & 32 ($P<5d$)      & \cite{MillerRicci2008b} \\
HD 209458 	&$70\pm50$ 	& 0.3 & --   & 20  & 0.3   & 0.3  & --                          &\cite{MillerRicci2008a} \\
	 		&			&  --   &  --  & --    & --      &   --   & 17 ($P<7d$),     &   \cite{Agol2007}\\
	 		&			&  	 &       &       &          &        & 100 ($P<10d$)  &  \\
TrES-1 		&$107\pm17$	& 1 	 & 2    & 3    & 2      & 1     & 100 ($P<6d$)    &\cite{Steffen2005,Rabus2009}\\
TrES-2 		&$257\pm27$	& 2 	 & 15  & 30  & --     & 1     & $50$ ($P<7d$)  &\cite{Rabus2009}\\
TrES-3 		&$70\pm30$	& 3-4 & --   & --     & --     & 10-15  &  --                    &\cite{Gibson2009a}  \\
\enddata
\tablenotetext{a}{A perturber in an n:m resonance completes $m$ orbits while the known planet completes $n$.}
\label{table:ttvlit}
\end{deluxetable}

\begin{deluxetable}{llllllll}
\tabletypesize{\scriptsize}
\tablecaption{Selected Parameters for OGLE-TR-111b}
\tablehead{Radius\tablenotemark{a}		& Midtime		& Duration	& Impact parameter & Filter	 & Reference  \\
($R_J$)			& (JD)						& (hours)			&				&			&}
\startdata
$0.97\pm0.06 $	& $24552330.44867	$ (fixed)	 	& --	&$0-0.68$		&\emph{I}	 	&  \cite{Pont2004, Santos2006}\\
$1.01\pm0.06$		& $24553470.56389 \pm 0.00055$	&$2.9\pm0.1$		&$0-0.65$		&\emph{V}	&  \cite{Minniti2007, Diaz2008} \\
$1.067\pm0.054$	& $24553787.70854 \pm 0.00035$	&$2.743\pm0.033$	&$0.25-0.55$		&\emph{I}		&  \cite{Winn2007} \\
$1.067\pm0.054$	& $24553799.75138 \pm 0.00032$	&$2.743\pm0.033$	&$0.25-0.55$		& \emph{I}		&  \cite{Winn2007}  \\
$0.922\pm0.057$\tablenotemark{b}	& $24554088.79145 \pm 0.00045$	&$2.67\pm0.014$	&$0.38\pm0.2$		& \emph{I}		&  \cite{Diaz2008}   \\
$0.922\pm0.057$\tablenotemark{b}	& $24554092.80493 \pm 0.00045$	&$2.67\pm0.014$	&$0.38\pm0.2$		& \emph{I} 	&  \cite{Diaz2008}   \\
\\							
$1.025\pm0.047$	& $24553470.56385 \pm 0.00100$	&$2.75\pm0.11$	&$0.39\pm0.28$	&\emph{V}	&  this work\tablenotemark{d}  \\
$1.066\pm0.038$	& $24553787.70860 \pm 0.00076$	&$2.73\pm0.047$	&$0.30\pm0.14$	&\emph{I}		&  this work\tablenotemark{d}   \\
$1.062\pm0.051$	& $24553799.75135 \pm 0.00090$	&$2.76\pm0.065$	&$0.39\pm0.16$	& \emph{I}		&  this work\tablenotemark{d}   \\
$0.952\pm0.076$	& $24554088.79118 \pm 0.00138$	&$2.62\pm0.060$	&$0.38\pm0.12$	& \emph{I}		&  this work\tablenotemark{d}   \\
$0.968\pm0.064$	& $24554092.80494 \pm 0.00129$	&$2.66\pm0.071$	&$0.35\pm0.15$	& \emph{I} 	&  this work\tablenotemark{d}   \\

$1.045\pm0.036$	& $24554574.54805 \pm 0.00078$	&$2.40\pm0.393$\tablenotemark{c}&$0.51\pm0.27$	& \emph{i'} 	&  this work\tablenotemark{d}   \\
$1.030\pm0.029$	& $24554578.55395 \pm 0.00061$	&$2.73\pm0.062$	&$0.36\pm0.19$	& \emph{i'} 	&  this work\tablenotemark{d}   \\
$0.996\pm0.022$	& $24554598.62680 \pm 0.00047$	&$2.63\pm0.037$	&$0.22\pm0.11$	& \emph{i'} 	&  this work\tablenotemark{d}   \\
$1.028\pm0.021$	& $24554602.64098 \pm 0.00053$	&$2.69\pm0.051$	&$0.30\pm0.12$	& \emph{i'} 	&  this work\tablenotemark{d}   \\
$1.006\pm0.025$	& $24554879.63787 \pm 0.00056$	&$2.68\pm0.054$	&$0.28\pm0.12$	& \emph{i'} 	&  this work\tablenotemark{d}   \\
$1.025\pm0.022$	& $24554903.72515 \pm 0.00039$	&$2.61\pm0.058$	&$0.16\pm0.11$	& \emph{i'} 	&  this work\tablenotemark{d}   \\
\enddata
\tablenotetext{a} {Except as noted, assuming $R_* = 0.83 R_{\odot}$}
\tablenotetext{b} {Using $R_* = 0.811 R_{\odot}$; if $R_* = 0.83 R_{\odot}$ is used, the radius is $R_P = 0.944 R_{J}$}
\tablenotetext{c} {Half-transit}
\tablenotetext{d} {All values from this work are taken from individual MCMC fits to each light curve, with with errors inflated to account for correlated noise; see \S~\ref{section:systematics}. }
\label{table:litparams}
\end{deluxetable}

\begin{deluxetable}{lllllllllll}
\tabletypesize{\scriptsize}
\tablecaption{Observational and Aperture Photometry Parameters for Six New Transits}
\tablehead{
Transit		& Frames 				& Exp. Time	& Binning	& Readout 	& $N_{Comp}$	& Aperture\tablenotemark{a}	& Sky radius	&Sky width & Scatter\tablenotemark{b} & Est. Poisson	\\
(UT)			& used (discarded)		& (sec)		& 		& (sec)		& 			&  (pixels)		& (pixels)		& (pixels)		& (mmag)	& (mmag)}
\startdata
20080418		& 178 (0)					& 30-60		& 1x1	& 5					& 3			& 12.8	& 35		&15		& 1.3		& 1.1		\\
20080422		& 110 (21\tablenotemark{c})	& 120		& 1x1	& 5					& 4			& 16.4	& 30		&10		& 2.0		& 0.8		\\
20080512		& 241 (20\tablenotemark{d})	& 60			& 1x1	& 5	 				& 3			& 19.2	& 30		&15		& 1.5		& 0.8		\\
20080516		& 276 (4\tablenotemark{e})	& 30-100		& 1x1	& 5					& 7			& 17.4	& 30		&10		& 1.5		& 0.8		\\
20090217		& 800 (0)					& 30			& 2x2	& 0.003\tablenotemark{f}	& 6			& 19.2	& 25		&20		& 1.2		& 1.1		\\
20090313		& 600 (42\tablenotemark{g})	& 15-30		& 2x2	& 0.003\tablenotemark{f}	& 1			&   9.6	& 20		&30		& 1.5		& 1.5		\\
\enddata
\tablenotetext{a} {Radius around star.}
\tablenotetext{b} {Standard deviation of the residuals on data binned to 120~s.}
\tablenotetext{c} {Insufficient counts on target.}
\tablenotetext{d} {Elongated images due to tracking failure.}
\tablenotetext{e} {Initial telescope focus not yet settled (3 points) and strongly aberrant ratio (1 point).}
\tablenotetext{f} {Frame transfer mode.}
\tablenotetext{g} {Comparison star saturated.}
\label{table:obsparams}
\end{deluxetable}

\clearpage

\begin{deluxetable}{l l l l }
\tablewidth{0pt}
\tabletypesize{\scriptsize}
\tablecaption{Flux Values For New Transits of OGLE-TR-111b \tablenotemark{a}}
\tablehead{ Mid-exposure (UTC)	& Mid-exposure (BJD)	& \textrm{Flux}  		& Error}
\startdata
2454574.520046	&	2454574.52375	&	0.9820188	&	0.00153 \\
2454574.522041	&	2454574.525745	&	0.9797899	&	0.00153 \\
2454574.522868	&	2454574.526573	&	0.9824249	&	0.00153 \\
2454574.523734	&	2454574.527439	&	0.9802172	&	0.00153 \\
2454574.524134	&	2454574.527838	&	0.9828871	&	0.00153 \\
\nodata
\enddata     
\tablenotetext{a} {Full table available online.}
\label{table:data}
\end{deluxetable}


\begin{deluxetable}{l c c  c }
\tablewidth{0pt}
\tabletypesize{\scriptsize}
\tablecaption{Transit Parameters Based on Joint Fit to Eleven Light Curves}
\tablehead{		& \textrm{Median value}  	& Formal Error\tablenotemark{a}	& \textrm{Adopted Error} \tablenotemark{b}}
\startdata
\emph{Fitted Parameters} \\
 $k$						& \bf{0.1261} 	& +0.0008, -0.0009	 	& \bf{+0.0010, -0.0011}  \\
 $a/R_*$			 		& \bf{12.3}		& 0.2		 			& \bf{0.2} 		 \\
 $i$				 		& \bf{88.3}		& +0.3, -0.2  			& \bf{+0.3, -0.2}  		\\
 $u_{1,i'}$					& \bf{0.32}		& 0.03				& \bf{0.04} 			 \\
 $u_{2,i'}$ 				& \bf{0.252}	& (fixed) 				& \nodata		 \\
  $u_{1,I}$					& \bf{0.30}	 	& 0.04 				&  \bf{0.05}	\\
 $u_{2,I}$					& \bf{0.2582}	& (fixed) 				& \nodata		 \\
 $u_{1,V}$				& \bf{0.6228} 	& (fixed) 				& \nodata		\\
 $u_{2,V}$				& \bf{0.1587}	& (fixed) 				& \nodata	 	\\
 \textrm{$T_C - 2453470$}	& \bf{0.56486} 	&  +0.00065,-0.00067	& \bf{+0.00098, -0.00089} 	\\
 \textrm{$T_C - 2453787$} 	& \bf{0.70934} 	&  +0.00043,-0.00044	& \bf{+0.00131, -0.00137} 	\\
 \textrm{$T_C - 2453799$} 	& \bf{0.75212} 	&  0.00044  			& \bf{+0.00117, -0.00129} 	\\
 \textrm{$T_C - 2454088$} 	& \bf{0.79197} 	&  0.00028  			& \bf{+0.00071, -0.00077}		\\
 \textrm{$T_C - 2454092$}	& \bf{0.80573} 	&  0.00032			& \bf{+0.00080, -0.00097} 	 \\
 \textrm{$T_C - 2454574$}	& \bf{0.54282} 	&  +0.00040,-0.00039	& \bf{+0.00103, -0.00068} 	\\
 \textrm{$T_C - 2454578$} 	& \bf{0.55469} 	&  +0.00044,-0.00045	& \bf{+0.00061, -0.00061}		\\
 \textrm{$T_C - 2454598$} 	& \bf{0.62752} 	&  0.00029  			& \bf{+0.00055, -0.00047}		\\
 \textrm{$T_C - 2454602$} 	& \bf{0.64170} 	&  0.00025  			& \bf{+0.00044, -0.00046} 	\\
 \textrm{$T_C - 2454879$}	& \bf{0.63864} 	&  0.00022  			& \bf{+0.00044, -0.00035}		\\
 \textrm{$T_C - 2454903$} 	& \bf{0.72565} 	&  0.00025  			& \bf{+0.00029, -0.00042}		 \\
\\  
\emph{Derived Parameters}\\
 $b$				 		& \bf{0.35}			& +0.04, -0.06	 	& \bf{+0.04, -0.06}   \\
 $T_{14}$ (sec)		 		& \bf{9647}		& +47, -48			& \bf{+57, -60}   \\
 $R_p$ ($R_J$)\tablenotemark{c}  	& \bf{1.019}	& 0.025 			& \bf{0.026}   \\
 $a$ (AU)\tablenotemark{c} 	 	& \bf{0.0473}	& +0.0015, -0.0014	& \bf{+0.0015, -0.0014}   \\
\enddata     
\tablenotetext{a} {Formal  $68.3\%$ credible interval from MCMC fit of all 11 light curves jointly.}
\tablenotetext{b} {Adopted error from residual permutation method, if greater than the formal error; see \S~\ref{section:systematics}.}
\tablenotetext{c} {Assuming $R_*=0.83 \pm 0.02 ~R_{\odot}$ \citep{Santos2006} and using $R_J=71,492 km$.}
\label{table:mcmc}
\end{deluxetable}

\begin{deluxetable}{l c l c l c l c l}
\tablewidth{0pt}
\tabletypesize{\scriptsize}
\tablecaption{Transit Parameters Based on Individual Fits to Each Light Curve}
\tablehead{
Transit &$k$\tablenotemark{a} & $f$\tablenotemark{b} &$T_{14}$\tablenotemark{a} & $f$\tablenotemark{b} &$a/R_*$\tablenotemark{a} & $f$\tablenotemark{b} &$i$\tablenotemark{a} & $f$\tablenotemark{b} }
\startdata
20050409 & $0.127 \pm 0.0047$ & 1.0 & $9901 \pm 381$ & \nodata	& $11.7 \pm 1.2$ & 1.1	& $88.1 \pm 1.5$ & 1.3 \\
20060221 & $0.132 \pm 0.0038$ & 2.0 & $9829 \pm 170$ & 1.2		& $12.2 \pm 0.7$ & 1.4	& $88.6 \pm 0.8$ & 1.1\\
20060305 & $0.132 \pm 0.0051$ & 3.2 & $9920 \pm 234$ & 1.6		& $11.8 \pm 0.9$ & 1.6	& $88.1 \pm 0.9$ & 1.4\\
20061219 & $0.118 \pm 0.0076$ & 3.1 & $9436 \pm 217$ & 1.0		& $12.2 \pm 0.8$ &  \nodata	& $88.2 \pm 0.9$ &  \nodata \\
20061223 & $0.120 \pm 0.0064$ & 1.8 & $9580 \pm 255$ & 1.2		& $12.2 \pm 0.9$ & 1.1	& $88.3 \pm 0.9$ &  \nodata \\
20080418 & $0.129 \pm 0.0036$ & 1.2 & $8646 \pm 1413$\tablenotemark{c} & 3.2	& $12.9 \pm 1.4$ & 1.8	& $87.7 \pm 1.5$ & 2.5\\
20080422 & $0.128 \pm 0.0029$ & 1.2 & $9837 \pm 222$ & \nodata		& $12.9 \pm 0.9$ & 1.2	& $88.3 \pm 1.0$ & 1.1 \\
20080512 & $0.123 \pm 0.0022$ & 1.9 & $9482 \pm 134$ & 1.5		& $12.8 \pm 0.6$ & 1.5	& $89.0 \pm 0.6$ & 1.0\\
20080516 & $0.127 \pm 0.0021$ & 1.5 & $9677 \pm 183$ & 1.4		& $12.3 \pm 0.7$ & 1.3	& $88.6 \pm 0.7$ &  \nodata \\
20090217 & $0.125 \pm 0.0025$ & 2.2 & $9644 \pm 193$ & 1.9		& $12.4 \pm 0.7$ & 1.6	& $88.7 \pm 0.7$ & 1.1\\
20090313 & $0.127 \pm 0.0022$ & 1.9 & $9390 \pm 209$ & 2.7		& $13.0 \pm 0.6$ & 2.2	& $89.2 \pm 0.6$ & 1.3\\
\enddata
\tablenotetext{a} {Formal individual MCMC fit value and error (scaled upward by factor $f$ in adjacent column).}
\tablenotetext{b} {Factor by which the error in the previous column has been increased based on the residual permutation method; no value is given if the formal MCMC fit error was larger. See \S~\ref{section:systematics}.}
\tablenotetext{c} {The ill-constrained duration of the half-transit 20080418 was not used in any fits.}
\label{table:indfits}
\end{deluxetable}

\begin{deluxetable}{l r r r r}
\tablewidth{0pt}
\tabletypesize{\scriptsize}
\tablecaption{Timing Residuals}
\tablehead{Transit & Number & O-C (s)&  $\sigma$}
\startdata
20050409\tablenotemark{a} &$-155$ 	& $-271 \pm 83 $ 	& $-3.3$ \\
20050409\tablenotemark{b} &$-155$ 	& $65 \pm 43$ 		& $1.5$ \\
20060221 	& $-76$ 	& $8 \pm 75$ 		& $0.1$ \\
20060305 	& $-73$ 	& $-41 \pm 82$ 	& $-0.5$ \\
20061219 	& $-1$	& $-66 \pm 103$ 	& $-0.6$ \\
20061223 	& $0$ 	& $-125 \pm 114$ 	& $-1.1$ \\
20080418 	& $120$	& $180 \pm  87$ 	& $2.1$ \\
20080422	 	& $121$ 	& $-42 \pm  52$ 	& $-0.8$ \\
20080512 	& $126$ 	& $10 \pm  41$ 	& $0.3$ \\
20080516 	& $127$	& $-13 \pm 46$ 	& $-0.3$ \\
20090217 	& $196$ 	& $-1 \pm 37$ 		& $-0.02$ \\
20090313 	& $202$ 	& $28 \pm 36$ 		& $0.8$ \\
\enddata
\tablenotetext{a} {Using our analysis of the original photometry from \citet{Minniti2007}.}
\tablenotetext{b} {Using the published time from reanalyzed photometry by \citet{Pietrukowicz2010}.}
\label{table:ominuscvalues}
\end{deluxetable}

\clearpage

\end{document}